\documentclass[aps,
prb,
reprint,
floatfix,
]{revtex4-2}
\usepackage{graphicx}
\usepackage{color}
\usepackage{xcolor}
\usepackage{amsmath}
\usepackage{amssymb}
\usepackage{setspace}
\usepackage{lineno}
\usepackage[normalem]{ulem}
\usepackage{bm}

\graphicspath{{figure/}}

\begin{document}

\title{Graph-Dynamics correspondence in metallic glass-forming liquids}

\author{Xin-Jia Zhou}
\author{Feng Yang}
\author{Xiao-Dong Yang}
\author{Lin Ma}
\author{Zhen-Wei Wu}
\email[]{zwwu@bnu.edu.cn}
\affiliation{Institute of Nonequilibrium Systems, School of Systems Science, Beijing Normal University, 100875 Beijing, China}

\date{\today}

\begin{abstract}

Theoretical challenges in understanding the nature of glass and the glass transition remain significant open questions in statistical and condensed matter physics. As a prototypical example of complex physical systems, glasses and the vitrification process have been central research topics, consistently attracting broad scientific interest. This focus has driven extensive studies on phenomena such as aging, non-exponential relaxation, dynamic anomalies, glass-forming ability, and the mechanical response of glasses under stress. Recent advances in computational and experimental techniques have enabled rigorous testing of theoretical models, shedding new light on glassy behavior. However, the intrinsic complexity of glass and the glass transition that lies in their physics, which spans multiple length and time scales, makes the system challenging to characterize. In this review, we emphasize the need to move beyond conventional approaches and propose a topological perspective as a promising alternative to address these challenges. Specifically, our findings reveal that the diversity in particle relaxation behavior is statistically linked to a global topological feature of the transient network structures formed by the particles in a given liquid. This direction offers opportunities to uncover novel phenomena that could fundamentally reshape our understanding of glassy materials.\\


\end{abstract}
\maketitle

\section{Introduction}

When a liquid is rapidly cooled below its melting point, it can bypass crystallization and form an amorphous solid known as glass~\cite{anderson1995through, berthier2011theoretical,Schroers2010precessing, Chen2011brief}. Although glass exhibits mechanical rigidity similar to that of crystalline solids, it lacks the long-range atomic order characteristic of crystals~\cite{kittel2018introduction, Kauzmann1948the}. This disordered state gives rise to unique and complex dynamic behaviors, including a dramatic slowdown in molecular motion during vitrification, with relaxation times spanning over 15 orders of magnitude~\cite{Ediger1996supercooled, Angell1991relaxation}. These dynamics are accompanied by distinctive phenomena, such as non-exponential relaxation, deviations from the Stokes-Einstein relation, and a non-Arrhenius temperature dependence of viscosity~\cite{debenedetti2001supercooled, Stillinger2013glass, Cao2020revisiting, Han2011transport, Chathoth2010stokes, Ren2021structural, Zhang2010fragile, Dyre2006colloqium}. Despite substantial progress in research, the microscopic nature of glass and the underlying mechanisms driving these behaviors remain unresolved, which presents one of the most challenging questions in solid-state physics~\cite{tanaka2010critical}.

To illustrate the complexity of glass, it is instructive to start with a familiar example: water. At room temperature, water exists as a disordered liquid, with atoms lacking a fixed arrangement. Upon cooling below its melting point, the water freezes into ice, adopting a crystalline structure that exhibits mechanical rigidity. In the framework of condensed matter physics, this rigidity arises from symmetry breaking, where a reduction in symmetry corresponds to the formation of an ordered phase~\cite{landau2013statistical}. However, amorphous materials such as glass do not follow this conventional route to rigidity. They exhibit characteristics reminiscent of both liquid water and solid ice: on the atomic scale, they remain disordered like a liquid, yet on a macroscopic scale, they possess the mechanical rigidity of a crystalline solid. Consequently, glasses are often defined by what they lack rather than what they possess, specifically, the absence of long-range order. This unconventional status underlines our limited understanding of the fundamental nature of glass and highlights why the glass transition remains a central and challenging issue in condensed matter physics~\cite{ediger2012perspective}.

\begin{figure}[ht]
\centering
\includegraphics[width=0.99\columnwidth]{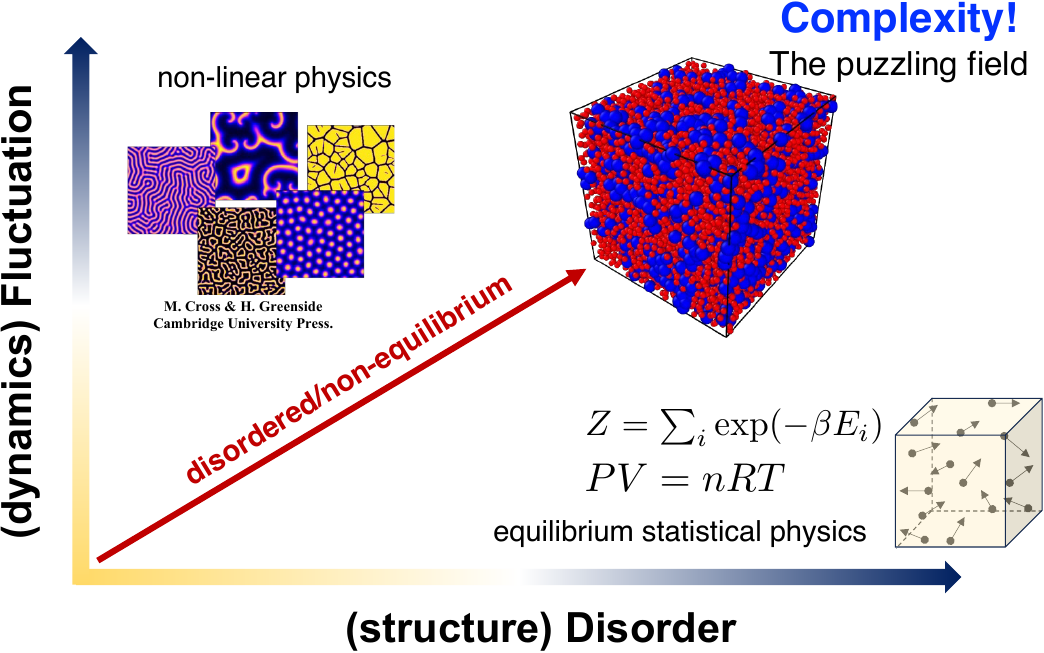}
\caption{{\bf The complexity nature of the study of glass physics.} Homogeneous systems exhibit regularities and symmetries that greatly facilitate understanding and analysis. However, in glass and other systems characterized by randomness and disorder, these simplifying regularities are absent. The inherent complexity of amorphous matter calls for innovative approaches and interdisciplinary collaboration to drive meaningful progress in the field.}
\label{fig1:complexity}
\end{figure}

Beyond their structural disorder, glasses also exhibit a diverse spectrum of dynamic behaviors~\cite{wang2019dynamic}, further complicating their characterization. Traditional methods in statistical mechanics often focus either on equilibrium systems with large numbers of degrees of freedom or on systems with relatively few degrees of freedom where non-equilibrium processes can be described by nonlinear equations~\cite{cross2009pattern}. In contrast, glasses encompass both extensive atomic-scale disorder and complex dynamic processes spanning a broad range of timescales. These processes interact intricately with the disordered atomic structure, making understanding the structure–dynamics relationship critical to developing a comprehensive theory of glasses (see Fig.~\ref{fig1:complexity}). In the following, we first address the structural aspects of glasses and then examine their connection to various dynamic processes, highlighting how these intertwined factors shape the behavior of amorphous materials.

Over the past half century, researchers have made significant progress in addressing these challenges. The developments range from the early concept of random close packing to the identification of critical atomic clusters through computer simulations, and more recently, the cluster packing model, which sheds light on medium-range order in amorphous materials~\cite{bernal1959geometrical,finney1970random,miracle2004structural,sheng2006atomic,zeng2011long,pan2011origin,Jiang2021non-monotonic}. Each advance has been closely linked to emerging concepts or technological innovations. Despite these breakthroughs, formidable challenges persist. Chief among these is the inherent complexity of the atomic structure in amorphous matter, making it difficult to reconcile experimental observations with theoretical predictions. Given that physics is fundamentally an experimental science, it is instructive to revisit what experiments have revealed. In particular, two studies have greatly influenced our research strategy. One showed a scale-invariant relationship between the molar atomic volume and the main peak position in the structure factor for nearly 40 metallic glasses, with a power-law exponent of 2.31, significantly different from the exponent of 3 observed in crystals~\cite{ma2009power}. Another investigation examined the reduced peak positions in the pair correlation function of more than 60 metallic glasses, uncovering a universal step-like sequence in real-space~\cite{liu2010metallic}. These scattering data exposed universal features of the atomic structure in glasses, suggesting a fundamental ``law of invariance''. Inspired by these findings, we have adopted a topological framework as a central approach to our research over the past decade~\cite{wu2013effect,wu2013correlation,wu2015hidden,wu2016critical,wu2018stretched,wu2020linking,wu2023topology,ma2024unveiling}.

Our approach to understanding the atomic structure of amorphous and supercooled liquids has shifted from concentrating solely on the variety and distribution of local atomic clusters to examining the connectivity among them~\cite{wu2013effect,wu2013correlation,wu2015hidden,wu2016critical,wu2018stretched,wu2020linking,ma2024unveiling}. In amorphous systems, the atomic structure is geometrically complex; for instance, Voronoi analysis often identifies thousands of distinct atomic clusters, each defined by its own local geometrical arrangement~\cite{Hirata2011direct}. This considerable diversity complicates standard structural analyses. Instead, by focusing on the local-linking topology among particles that share the same local symmetry, we gain a more transparent and tractable perspective on the relationship between structure and dynamics. Building on this topological framework, we have conducted a series of investigations into how specific connections between symmetrically similar particles influence their relaxation dynamics at the particle level. Drawing inspiration from the concept of ``node degree'' in complex network and graph theory~\cite{newman2003structure,Zhan2010on}, we introduced a straightforward structural order parameter termed {\it local connectivity}.
This parameter, analogous to counting the number of nearest-neighbor spins aligned with a central spin in an Ising model, has demonstrated clear correlations with various physical properties, in which includes relaxation dynamics~\cite{wu2013correlation,wu2018stretched,wu2020linking}, transport behavior~\cite{wu2018stretched,wu2020linking}, local atomic-cluster symmetry~\cite{wu2016critical,wu2020linking,ma2024unveiling}, and the thermal stability of supercooled liquids~\cite{desgranges2018unusual,desgranges2019can,ma2024unveiling}.
Collectively, these findings offered valuable insights into key characteristics of glasses and glass-forming liquids, such as non-exponential relaxation, super-Arrhenius behavior, and the percolation processes underlying vitrification.

Our paper is organized as follows. In Section~\ref{sec2}, we provide a concise discussion of the intermediate scattering function, which forms the basis of our dynamical analysis. In Section~\ref{sec3}, we introduce the graph-representation method and offer a detailed summary of the relationship between local connectivity and particle-level dynamics in metallic glass-forming liquids. This section concludes with a discussion on why local connectivity governs a wide range of physical properties in these systems. Section~\ref{sec4} presents our new results from a parallel comparative study of CuZr and NiAl. Here, we demonstrate that the diversity in particle relaxation behavior is statistically linked to a global topological feature of the transient network formed by the particles in a given liquid, leading us to introduce the concept of {\it graph-dynamics correspondence}. Finally, Section~\ref{sec5} outlines our perspective and outlook on open questions and potential future research directions.

\section{A presentation of the Intermediate Scattering Function}\label{sec2}

Here, we begin by introducing the intermediate scattering function $F(q,t)$, a crucial observable that is important from both the theoretical and experimental points of view, since it can be measured experimentally. It is a space-time correlator related to the Fourier-space density-density correlation function~\cite{kob1995testing} and contains all information about the relaxation dynamics in the system on the spatial scale of $1/q$,
\begin{equation}
    F({\boldsymbol q},t)=\frac{1}{N} \left \langle \sum_{ij} \exp \{{\rm i} {\boldsymbol q}\cdot [{\boldsymbol r}_j(t)-{\boldsymbol r}_i(0)] \} \right \rangle~,
\end{equation}
and when $i=j$, this function simplifies to the self-intermediate scattering function (SISF), which is computationally efficient as it excludes cross-terms. At large $q$ values, omitting cross-terms has minimal impact on the result; however, at small $q$ values, this distinction becomes significant. This is because, at smaller $q$, the scattering function captures information over larger spatial scales, where the collective effects (many bodies) between particles are more pronounced~\cite{hansen2013theory,binder2011glassy}.
Consider a simple example~\cite{hansen2013theory} where the motion of the particles follows the diffusion equation $\partial_t \rho^{\rm s}_{\boldsymbol q}(t) = -Dq^2\rho^{\rm s}_{\boldsymbol q}(t)$. If ones substitute its solution in wavevector space $\rho^{\rm s}_{\boldsymbol q}(t) = \rho^{\rm s}_{\boldsymbol q}\exp(-Dq^2t)$ into the definition of the self-intermediate scattering function,
\begin{equation}
\resizebox{.905\hsize}{!}{$F_{\rm s}({\boldsymbol q},t)=\frac{1}{N} \left \langle \rho^{\rm s}_{-\boldsymbol q}(0)\rho^{\rm s}_{\boldsymbol q}(t) \right \rangle = \frac{1}{N} \left \langle \rho^{\rm s}_{-\boldsymbol q}\rho^{\rm s}_{\boldsymbol q} \right \rangle \exp(-Dq^2t) =\exp(-Dq^2t)$~,}
\end{equation}
then ones obtain a straightforward relaxation behavior: an exponential decay. The exponent's prefactor depends on both the wave-vector $q$ and the diffusion coefficient $D$, allowing ones to experimentally measure the diffusion coefficient of the system from these scattering data.

\begin{figure}[ht]
    \centering
    \includegraphics[width=0.85\columnwidth]{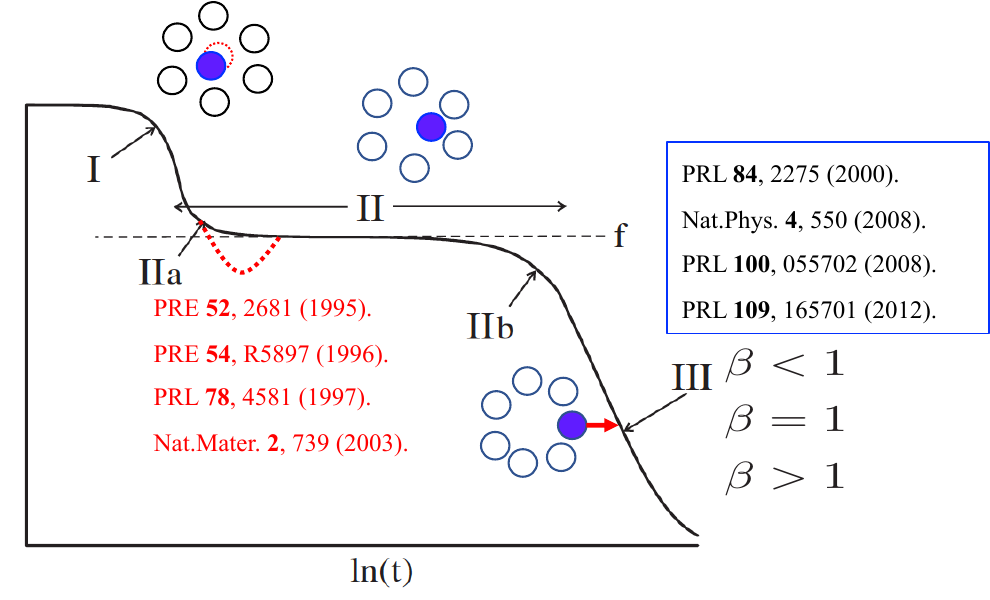}
    \caption{{\bf Typical intermediate scattering function of a supercooled liquid as a function of time (adapted from~\cite{reichman2005mode}).} Rather than following a simple exponential decay, the function exhibits multiple temporal regimes, as detailed in the text. The data are plotted on a logarithmic scale to clearly highlight the distinctions among these regimes. Note that the origin of the ``bump'' signal (red arc) can vary significantly between different systems; please refer to the listed literature for further details.
    }
    \label{fig2:isfex1}
\end{figure}

\begin{figure}[ht]
    \centering
    \includegraphics[width=0.75\columnwidth]{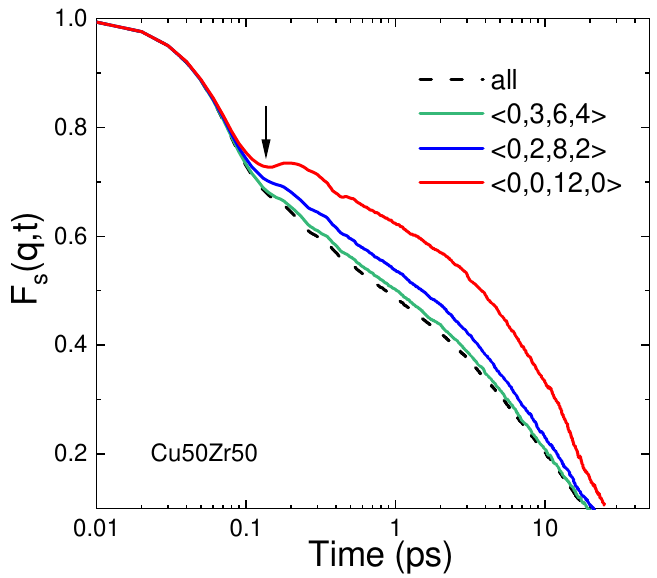}
    \caption{{\bf A more specific case: CuZr metallic glass-forming liquids.} Self-intermediate scattering function for $q=2.8$\AA$^{-1}$ at $T=1100$K. The different curves correspond to particles with different local coordinated arrangements characterized by the corresponding index of Voronoi analysis~\cite{finney1970random}. Also included is the data for all atoms.}
    \label{fig3:isfex2}
\end{figure}

However, in typical supercooled glass-forming liquids, the situation is much more complex~\cite{reichman2005mode}. As illustrated in Fig.~\ref{fig2:isfex1}, the decay forms of the relaxation curves vary on different timescales, each corresponding to distinct physical processes~\cite{ruta2017relaxation}. In the sub-picosecond regime (I), atomic dynamics stem from free motion and collisions, causing the ISF to decay rapidly from its initial value of 1. At later times, both self- and collective particle motion can often be described by mode-coupling theory (MCT), which has successfully characterized the ISF in a variety of simple and glass-forming liquids at temperatures $T > T_\mathrm{c}$, where $T_\mathrm{c}$ denotes the MCT crossover temperature~\cite{gotze2009complex,gotze1992relaxation}. As atomic packing fractions increase or temperatures drop, structural arrest sets in, producing a long-time plateau (regime-II) in the ISF defined by $f(q,T)$, also referred to as the Debye–Waller factor or the non-ergodicity parameter, as illustrated in Fig.~\ref{fig2:isfex1}. From a microscopic perspective, this plateau can be viewed as atoms ``rattling'' within transient local cages formed by their nearest neighbors. Such cage dynamics become especially pronounced in viscous liquids approaching $T_\mathrm{c}$~\cite{gotze1992relaxation,teichler2005heterogeneous}. In some cases, the relaxation curve exhibits special features, such as the damping valley illustrated by the red arc on short time scales in Fig.~\ref{fig2:isfex1}, which often indicates unique underlying physical processes and is considered a typical feature of strong liquids~\cite{angell1995formation,horbach1996finite,kob1997aging,horbach2001high,sastry2003liquid} in the sense of fragility. Since this local minimum is thought to be associated with the boson peak~\cite{sastry2003liquid,Habasaki1995origins}, it is important to carefully distinguish between finite-size effects and genuine dynamic features of strong glass formers. In long-times relaxation regime, the decay of the function could be fitted to a stretched exponential law $F(t)=\exp[-(t/\tau)^{\beta}]$ with $0<\beta<1$ in general, and $\beta$, $\tau$ will be wavevector-$q$ and temperature dependent. Interestingly, on long timescales, the shape factor $\beta$ of the relaxation curve may change to a value larger than $1$~\cite{ruta2012atomic,luo2017relaxation,chen2024understanding,bi2018multiscale}, suggesting the presence of unusual transport behaviors.
In Fig.~\ref{fig3:isfex2}, we present a case study of SISF for supercooled metallic glass-forming liquids at 1100 K. The overall signal (black dashed line) exhibits the typical relaxation characteristics of supercooled liquids. Furthermore, molecular dynamics simulations allow us to track the dynamics of atoms with distinct local symmetries, uncovering clear differences at the atomic level. Notably, icosahedral (ico) atoms show unique behavior, characterized by pronounced short-time damping and distinct long-time relaxation scales. In our {\it graph}-theoretical framework, these icosahedral atoms constitute the ``vertices''.

\section{Graph-ensemble and Relaxation dynamics of ico-atoms}\label{sec3}

\subsection{From MD-generated configurations to Graph-ensemble}

\begin{figure*}[ht]
    \centering
    \includegraphics[width=0.97\textwidth]{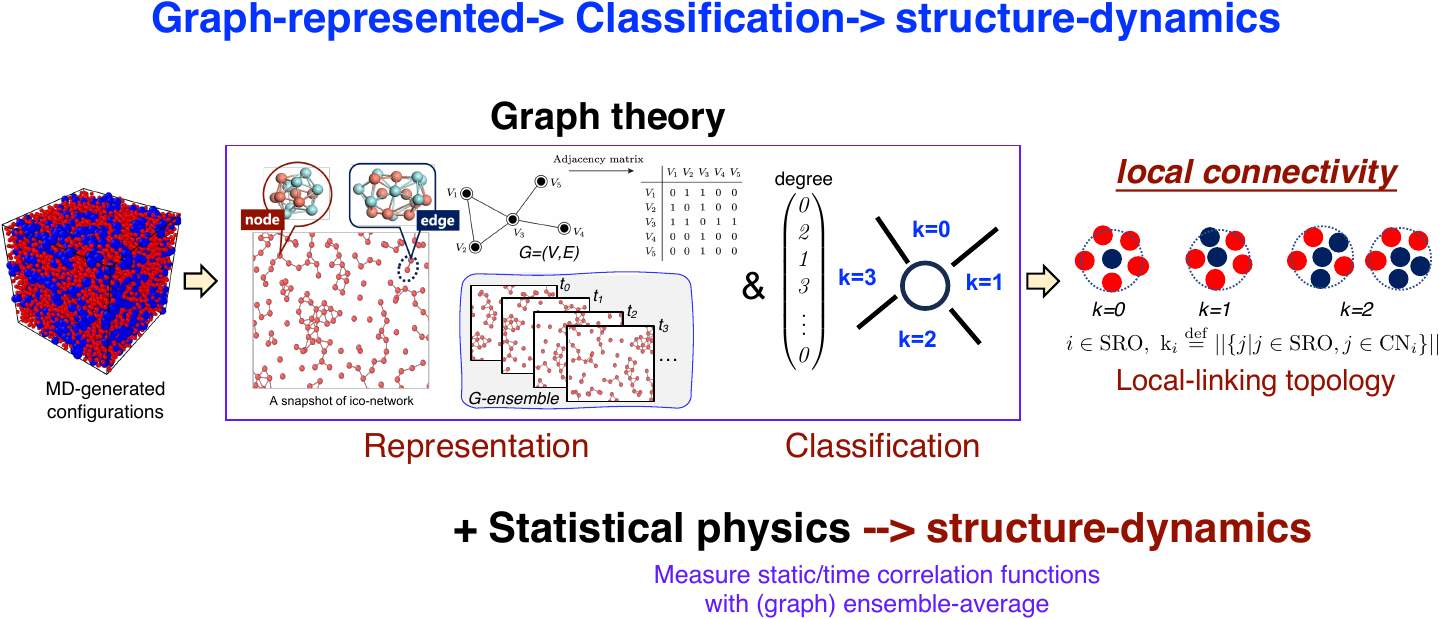}
    \caption{{\bf Graph-represented structures of disordered system.} A schematic diagram. {\it G}-ensemble is a graph-representation of an ensemble of MD-generated configurations. And here we are not dealing with dynamics on a graph with static topology but rather with ensemble-averaged physical dynamics that reflect realistic, evolving liquid-structures. {\it Degree}: In graph theory, the degree of a vertex is the number of edges connected to it. Although, in some cases, multiple edges may exist between two vertices, here we restrict our set to undirected, unweighted graphs, where the degree corresponds directly to the number of adjacent vertices. (Adapted from Ref.~\cite{wu2016critical,wu2018stretched,wu2020linking})
    }
    \label{fig4:graphtodyan}
\end{figure*}

\noindent
{\it Graph-representation and Classification---}
The core of this approach is mapping the atomic structure of supercooled liquids in graph-theoretic terms, in which the {\it vertex} and {\it edge} must be carefully defined. Once represented as a graph $G=(V,E)$, with $|V|$ nodes (vertices) and $|E|$ connections (edges), graph theory enables the classification of atoms solely by their local-linking topology~\cite{newman2003structure}, substantially simplifying the depiction of local atomic environments. The connections among network units are often encoded in an adjacency matrix $\mathbf{A}$, where $A_{ij} = 1$ if nodes $i$ and $j$ are connected and $A_{ij} = 0$ otherwise~\cite{wu2020linking}, i.e., the matrix elements indicate whether pairs of vertices share an edge (Fig.~\ref{fig4:graphtodyan}). The choice of ``bonds'' is thus crucial to ensuring a meaningful graph representation. Within this framework, one can calculate particle-level correlation functions, offering deeper insight into dynamic behaviors. Note that we do not analyze processes on a static graph; instead, we consider ensemble-averaged physical dynamics that mirror realistic, evolving liquid structures.

\vspace{5mm}

\subsection{Long-time relaxation and transport behaviors}

As discussed above in Fig.~\ref{fig3:isfex2}, in metallic liquid ico atoms constitute slowly relaxing structures, and hence we focus here on the dynamics of these entities. For this we use a Voronoi construction~\cite{finney1970random} to identify those Cu atoms that are in the center of an icosahedron. For each of these atoms we count the number, $k$, of neighboring Cu atoms that are also an icosahedral center and we refer to the central Cu atom as the particle with a degree of local connectivity equal to $k$. Here we consider the $k$-dependence of $F_{\rm s}(q,t)$. For this we investigated the following generalization of the self-intermediate scattering function:
\begin{equation}
\resizebox{.905\hsize}{!}{$F_{\rm s}({\boldsymbol q},k,t) = \left \langle \frac{1}{N_k(0)} \sum_{j=1}^{N_{\rm ico}} \exp \{{\rm i}{\boldsymbol q}\cdot [{\boldsymbol r}_j(t)-{\boldsymbol r}_j(0)] \} \delta[k_j(0)-k] \right \rangle$~,}
\end{equation}
where the $\delta$-function on the right-hand side makes sure that one considers only particles that at time $t=0$ were of that with a degree equal to $k$.

\begin{figure}[ht]
    \centering
    \includegraphics[width=0.99\columnwidth]{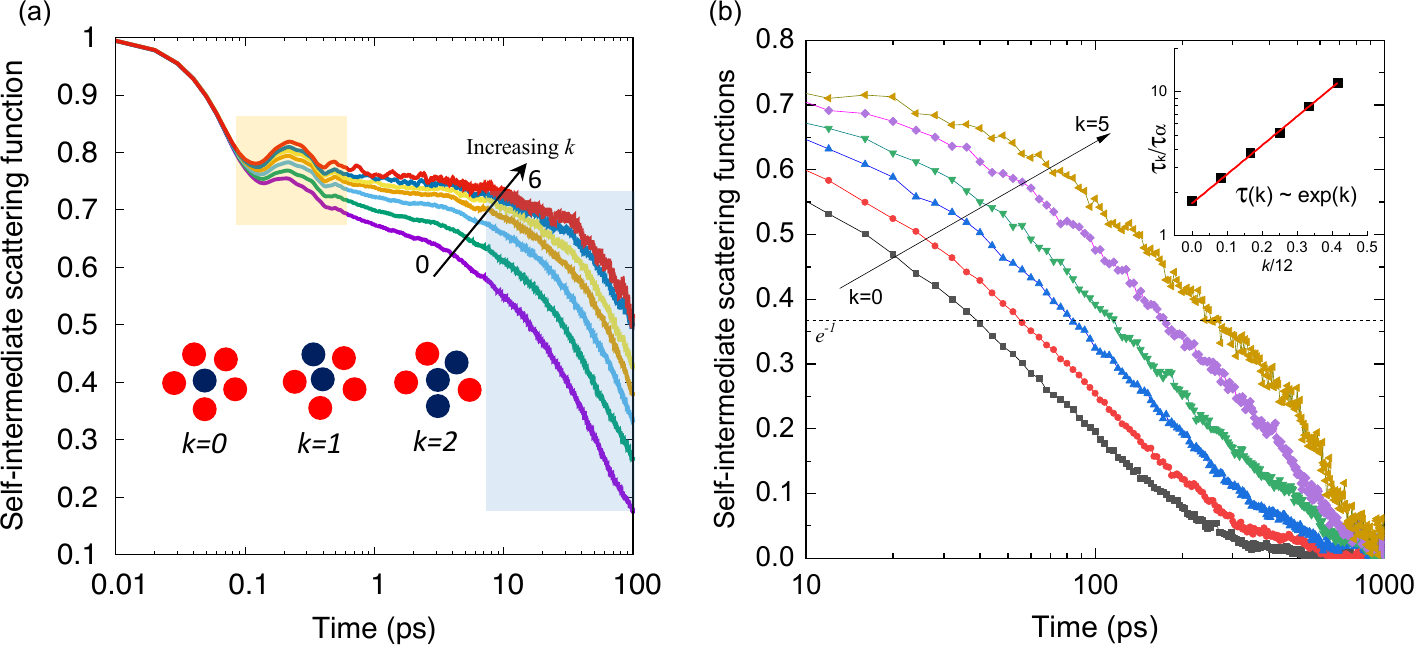}
    \caption{{\bf Time dependence of $F_{\rm s}(q,t)$ for particles that have different local connectivity $k$.} The wave-vector is 2.8\AA$^{-1}$ which corresponds to the main peak in the static structure factor and $T=1000$K. It has been recognized in Ref.~\cite{wu2018stretched} that this peak show basically no $T$-dependence. The main peak for the Cu-Cu correlation is around 2.8\AA$^{-1}$, the wave-vector we often focus on in our studies. The inset in panel (a) illustrates the definition of particles with different local connectivities $k$: Particles in blue are the center of an icosahedral-like cluster. The straight line in the upper inset of panel (b) is a exponential fit to the data of relaxation time $\tau$. (Adapted from Ref.~\cite{wu2018stretched,wu2016critical})
    }
    \label{fig5:isf_k}
\end{figure}

The results indicate that the dynamical relaxation of ico atoms in supercooled liquids strongly depends on their local-linking topology. As illustrated in Fig.~\ref{fig5:isf_k}(a), both the short- and long-timescale relaxation behaviors of these ico atoms are influenced by the connectivity of their surrounding atomic structures. Focusing on the long-time behavior, in Fig.~\ref{fig5:isf_k}(b) one observes that whether an ico atom relaxes more quickly or more slowly is directly related to its connectivity $k$. Atoms with higher $k$ values exhibit slower dynamics, with their relaxation times reaching an order of magnitude longer than the system averaged one. Considering the relaxation time over a range of $k$ values, defined as the time at which the relaxation function decays to $e^{-1}$, one finds a clear exponential relationship between the relaxation time $\tau$ and the connectivity $k$. This result suggests that local connectivity offers a quantitative structure–dynamics link in supercooled liquids, providing a simpler and more direct framework than typical map-to-map comparing approaches.

\begin{figure*}[ht]
    \centering
    \includegraphics[width=0.85\textwidth]{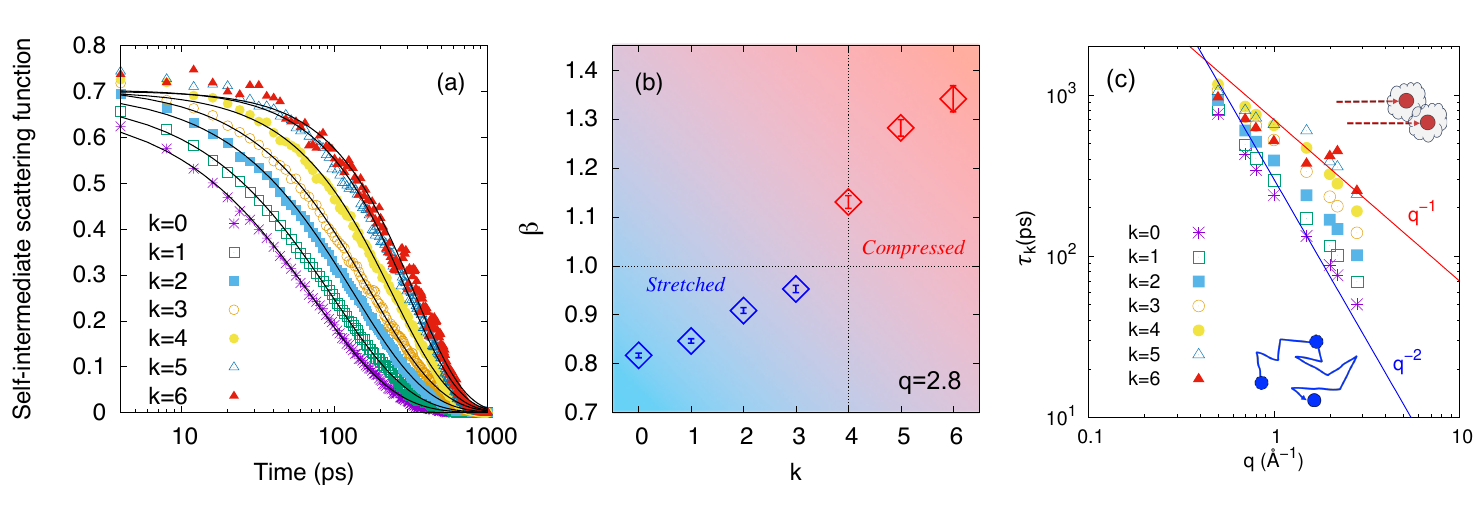}
    \caption{{\bf $k-$dependent transport behavior.} (a) $F_{\rm s}(q,t)$ at wave-vector $q=2.8$\AA$^{-1}$ for different values of connectivity $k$ of the Cu atoms. $T=1000$K. The solid lines are fits to the data with the KWW equation. (b) $k$-dependence of the KWW exponent $\beta$ at $q=2.8$\AA$^{-1}$. (c) $q$-dependence of relaxation time $\tau$ for particles with different values of $k$. The lower and upper two insets in panel (c) illustrate diffusive (Particles are scattered) and ballistic-like (Behaved like no scattering anymore) motion of particles, respectively. (Adapted from Ref.~\cite{wu2018stretched})
    }
    \label{fig6:isf_transport}
\end{figure*}

In addition to relaxation kinetics, the shape of the relaxation curve also depends on the connectivity of the particle. We fitted the SISF for populations with different $k$ with a Kohlrausch-Williams-Watts (KWW) function, i.e.~$F_{\rm s}(q,t)= A \exp(-(t/\tau)^\beta)$, where the prefactor $A$, the relaxation time $\tau$ and the shape factor $\beta$ depend on $q$. Figure~\ref{fig6:isf_transport}(a) displays this data and the fits and one sees that the correlator has in the $\alpha-$regime a very strong $k-$dependence. Figure~\ref{fig6:isf_transport}(b) presents the $k-$dependence of the KWW exponent $\beta$ for the wave-vector $q=2.8$\AA$^{-1}$. One recognizes that for small $k$, i.e.~isolated icosahedral clusters, $\beta$ is smaller than 1.0, i.e.~the relaxation is stretched as expected for a glass-forming system~\cite{binder2011glassy}. Interestingly, we find that for intermediate and large $k$ ($k\geq4$) the exponent increases significantly and becomes larger than 1.0, i.e.~one sees a crossover from normal glassy dynamics to one in which the correlator has a much sharper decay in time. Since vertices with $k=4$ (local tetrahedral-like building blocks) enhance the construction of a $3$-dimensional network, this dynamic transition may also imply some geometrical implications. It is noted that this crossover strongly depends on $q$~\cite{wu2018stretched}, i.e.~the length scale considered. The quick decay of the correlation function for large $k$ indicates a sudden yielding of the structure, i.e.~a type of motion that is very different from the viscous flow found in glassy systems. This crossover in dynamics is also reflected by the wave-vector dependence of the structural relaxation time $\tau$ (determined by the decays of $F_{\rm s}(q,t)$ to $e^{-1}$). In Fig.~\ref{fig6:isf_transport}(c), we observed that as the connectivity $k$ increases, the scaling law between the relaxation time and the wave-vector shifts from $q^{-2}$ to $q^{-1}$, which indicates that the relaxation processes within supercooled liquids are highly complex. For certain atoms coordinated by special local ordering, with the influence of their neighbors and much more flexible surroundings, they can have ballistic-like transport behavior. The results show that even in the fluid state certain glassy systems can show a relaxation dynamics that is given by a compressed exponential, i.e., a time dependence that differs strongly from the usual stretched exponential found in viscous liquids. For the metallic glass-former investigated here these two types of relaxation dynamics do coexist and can be directly related to the local atomic structure.

Although the different relaxation behavior is most easily seen if one considers structural entities that are beyond the atomic distances, a detailed analysis of the wave-vector dependence of the relaxation times of the particle-averaged $F_{\rm s}(q,t)$ does reveal the fingerprint of the two relaxation mechanisms. It can be hypothesized that the structural attributes underlying these mechanisms in the liquid phase may also influence the unique mechanical properties of metallic glasses, such as elasticity and ductility. Specifically, the more rigid networks formed by high-$k$ icosahedra appear to be embedded in a softer matrix, thereby facilitating localized plastic relaxation. A comprehensive investigation into how these structural relaxation processes correlate with mechanical behavior in metallic glasses~\cite{fan2024topology} could offer significant insight into the design and optimization of these materials.

\begin{figure}[ht]
    \centering
    \includegraphics[width=0.75\columnwidth]{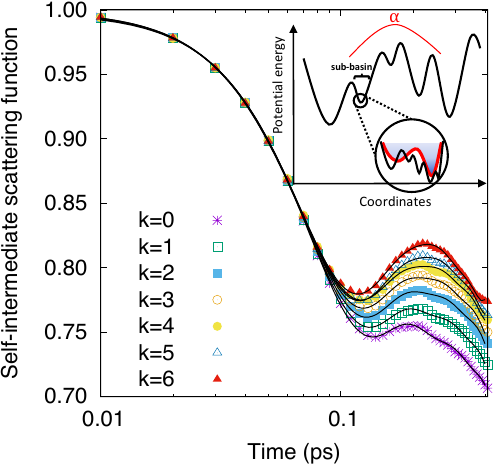}
    \caption{{\bf Relaxation dynamics of the ico-atoms at short times.} Short time behavior of the $F_{\rm s}(q,t)$ of particles with different local connectivity $k$ (symbols). The wave-vector is $q=2.8$\AA$^{-1}$ and $T=1000$K. With increasing $k$ the height of the peak at around 0.2~ps increases showing that the motion becomes less damped. The solid lines are fits to the data with Eq.~(\ref{eq:fsqfit}). The upper inset is an illustration of our fitting Ansatz of Eq.~(\ref{eq:fsqfit}) in a perspective of energy-landscape (Adapted from Ref.~\cite{wu2018stretched,wu2016critical})
    }
    \label{fig7:icodamping}
\end{figure}

\begin{figure}[ht]
    \centering
    \includegraphics[width=0.95\columnwidth]{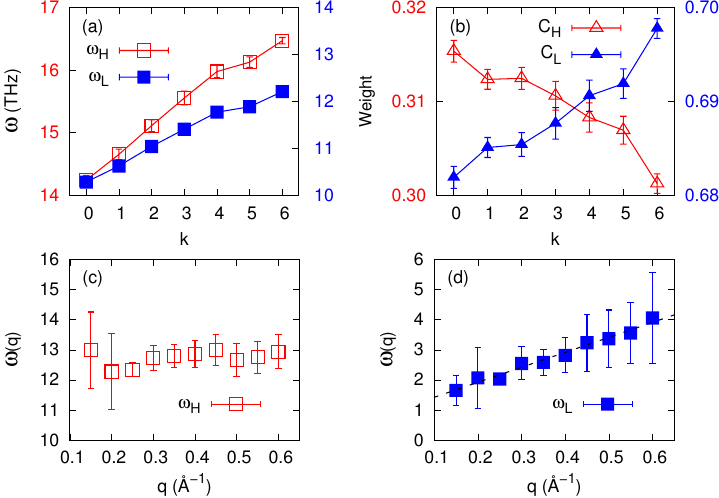}
    \caption{{\bf The dependence of dynamic features on connectivity $k$ and dispersion relations.} (a) Both the high and low frequency modes, $\omega_{\rm H}$ and $\omega_{\rm L}$, increases with increasing $k$. $q=2.8$\AA$^{-1}$. (b) The fraction of motion $C_{\rm H/L}$ increases for $\omega_{\rm L}$ and decreases for $\omega_{\rm H}$. (c) The high frequency mode $\omega_{\rm H}(q)$ is approximately $q-$independent, characteristic of localization of the vibrational modes. (d) The low frequency mode $\omega_{\rm L}(q)$ increases monotonically with increasing $q$, characteristic of collective motions. Error bars in these panels have been obtained from the fit of the scattering function with Eq.~(\ref{eq:fsqfit}). (Adapted from Ref.~\cite{wu2018stretched,wu2020linking})
    }
    \label{fig8:lcho}
\end{figure}

\subsection{Short-time damping valley}

To describe the dynamics at short-time scales in a quantitative manner, it was proposed that the correlator can be fitted with a simple Ansatz~\cite{wu2018stretched}:
\begin{footnotesize}
\begin{equation}
    F_{\rm s}(q,t)= \sum_{m={L,H}} C_m\exp\{{\rm i} q [A\cos(\omega_m t+\delta_m)-A\cos(\delta_m)]\}~,
\label{eq:fsqfit}
\end{equation}
\end{footnotesize}
\noindent
where $L$ and $H$ denote a low- and high-frequency mode, respectively, and we have $C_L+C_H=1$, i.e., this model proposes that collective particle motions ($\sim \sum_{j=1}^N [\cdots]$) can be represented by two harmonic oscillators ($\sim \sum C_L[\cdots] + C_H[\cdots]$) with distinct frequencies that dephase after an initial period of time evolution. The resulting fits are included in Fig.~\ref{fig7:icodamping} as well, and we see that this functional form gives a good description of the data. Figure~\ref{fig8:lcho}(a) presents the $k-$dependence of $\omega_H$ and $\omega_L$ and one recognizes that both increase (basically linearly) with $k$, demonstrating that with increasing connectivity the cage becomes stiffer. This outcome even allows for a minor, yet noteworthy, prediction. By combining the linear relationship between $k$ and $\omega$ with the standard textbook knowledge of the connection between spring constants and vibrational frequencies, it can be deduced that there is a quadratic relationship between local connectivity $k$ and elastic moduli at the atomic level in the system studied. In fact, a quadratic fit will align well with previous computational results~\cite{wakeda2010icosahedral}.
The $k$-dependence of the weighted parameters $C_{L/H}$, presented in Fig.~\ref{fig8:lcho}(b), shows different behaviors as $k$ increases. The fraction of the $\omega_L$ mode increases, while the fraction of the $\omega_H$ mode decreases. For the study of the collective properties of the icosahedral coordinated particels, the coherent ISF for all these special particles in a $q$-range of 0.15 \AA$^{-1}$ $\le q\le$ 0.6 \r{A}$^{-1}$ was also measured, and then the corresponding ISF was fitted by the proposed Ansatz. The obtained $\omega_H(q)$ and $\omega_L(q)$, are shown in Fig.~\ref{fig8:lcho}(c) and \ref{fig8:lcho}(d), respectively. It can be seen that $\omega_H(q)$ is approximately $q$-independent in the entire investigated range of $q$, indicating that the $\omega_H(q)$ mode is localized, suggesting an optical-like mode.
For example, if the center of an icosahedron vibrates in one direction while its ``shell'' moves in the opposite direction, this would represent a typical optical mode, resulting in no $q$-dependence. The so-called ``Einstein frequency'' for this system can be calculated as $\Omega_0^2 = \langle|\vec{F}_i|^2\rangle / 3mk_BT$, where $\langle\cdot\rangle$ denotes the average force acting on icosahedrally coordinated Cu atoms. Depending on the choice of $m$, i.e.~whether it represents the mass of the central Cu atom, all shell atoms, only the Cu atoms in the shell, only the Zr atoms in the shell, or the reduced mass of the core-shell system, the resulting $\Omega_0$ values are 29.25 THz, 7.57 THz, 12.7 THz, 9.43 THz, and 30.21 THz, respectively. Notably, the result for Cu atoms only in the shell closely matches the fitted frequency shown in Fig.~\ref{fig8:lcho}(c). In contrast, $\omega_L(q)$ shows a remarkable $q$-dependence, increasing linearly as $q$ increases at low $q$. This trend indicates that collective dynamics persist at these frequencies, as expected for a generic ISF that couples at low $q$ to the acoustic modes~\cite{sette1998dynamics}. The linear dispersion slope for $\omega_L(q)$ is approximately 500 m/s, closely matching the system's thermal velocity. This velocity can also be estimated using the prefactor $\omega_0^2 = q^2 k_B T / 2mS(q)$ from the $t^2$-term in the short-time Taylor expansion of the coherent ISF~\cite{hansen2013theory,binder2011glassy}. Accordingly, the analytic slope derived from $\sqrt{k_B T / 2mS(q)}$ is about 580 m/s, which aligns well with the linear fit of the data shown in Fig.~\ref{fig8:lcho}(d).

\vspace{5mm}

\noindent
{\it Summary and Discussion---}
Compressed-exponential behavior in correlation functions frequently appears in out-of-equilibrium systems, where hyperdiffusive motion dominates the dynamics. In glasses, this hyperdiffusion is attributed to the release of internal stress accumulated during quenching~\cite{cipelletti2000universal}. For nanoparticles suspended in a supercooled solvent, faster-than-exponential relaxations are hypothesized to arise from cooperative behavior prompted by the near-vitreous medium~\cite{caronna2008dynamics,guo2009nanoparticle}. In our study, both cooperative motion, i.e. involving the convection of stiffer units by more flexible surroundings, and internal stress release drive hyperdiffusive dynamics. First, the development of cooperative particle motions is evidenced by the positive correlation between local connectivity and the extended-mode contribution, as shown in Fig.~\ref{fig8:lcho}(b). Second, previous studies suggest that icosahedrally coordinated particles exhibit a higher average elastic modulus yet smaller atomic volume under increased local connectivity~\cite{wakeda2010icosahedral}, and consequently particles with higher connectivity store more localized stress. Therefore, hyperdiffusive motion in these highly connected particles also reflects the ongoing release of localized stress, akin to mechanisms proposed for colloidal gels, where local stress arises from elastic network deformation triggered by gel syneresis~\cite{cipelletti2000universal}.

These findings suggest that even in a simple metallic supercooled liquid, particles coordinated by specific spatial symmetries can exhibit unexpectedly complex dynamics. Molecular dynamics simulations have shown that annealing at temperatures just below the glass transition temperature increases the fraction of short-range icosahedral order in metallic glasses~\cite{zhang2014effects}. Due to a self-aggregation effect~\cite{li2009structural}, a higher proportion of icosahedra indicates an increase in particles with greater local connectivity, whose structural relaxation follows a compressed exponential decay. This observation is strikingly similar to recent experimental findings~\cite{luo2017relaxation} showing that pre-annealed metallic glasses at 0.9$T_{\rm g}$ exhibit fast dynamic modes ($\beta >$ 1) with ballistic-like behavior. The results reviewed here may offer a possible structural basis for this unusual dynamic behavior in metallic glasses.

Furthermore, it is well established that the intensity of the boson peak increases as the fragility of a glass-forming liquid decreases~\cite{angell1995formation,shintani2008universal}, where fragility characterizes how sharply the viscosity rises upon cooling. An intriguing open question is why this high-frequency feature, captured by the boson peak, correlates with the slowest flow dynamics determined by fragility. According to Angell's definition, network-like liquids (e.g., silica, which relies on strong directional covalent bonds) are considered \textit{strong} glass-formers~\cite{shintani2008universal,sastry2003liquid}. Consistent with the observed link between ``bond''-forming tendencies and fragility~\cite{angell1995formation,shintani2008universal}, our findings suggest a potential structural mechanism that bridges dynamic processes across multiple timescales. Specifically, as the number of particles with connectivity $k \geq 4$ increases, the resultant tetrahedral-like network (e.g., water with a fully developed hydrogen-bonding network) generally lowers the system's fragility~\cite{sastry2003liquid,liu2005pressure,soper2000structures,xu2005relation,lad2012signatures}. Simultaneously, particles with higher local connectivity contribute more significantly to the extended modes associated with boson peak dynamics. Consequently, an enhanced boson peak intensity is likely to accompany the growth of such network-like structures in supercooled liquids.

\vspace{5mm}

\begin{figure}[ht]
    \centering
    \includegraphics[width=0.95\columnwidth]{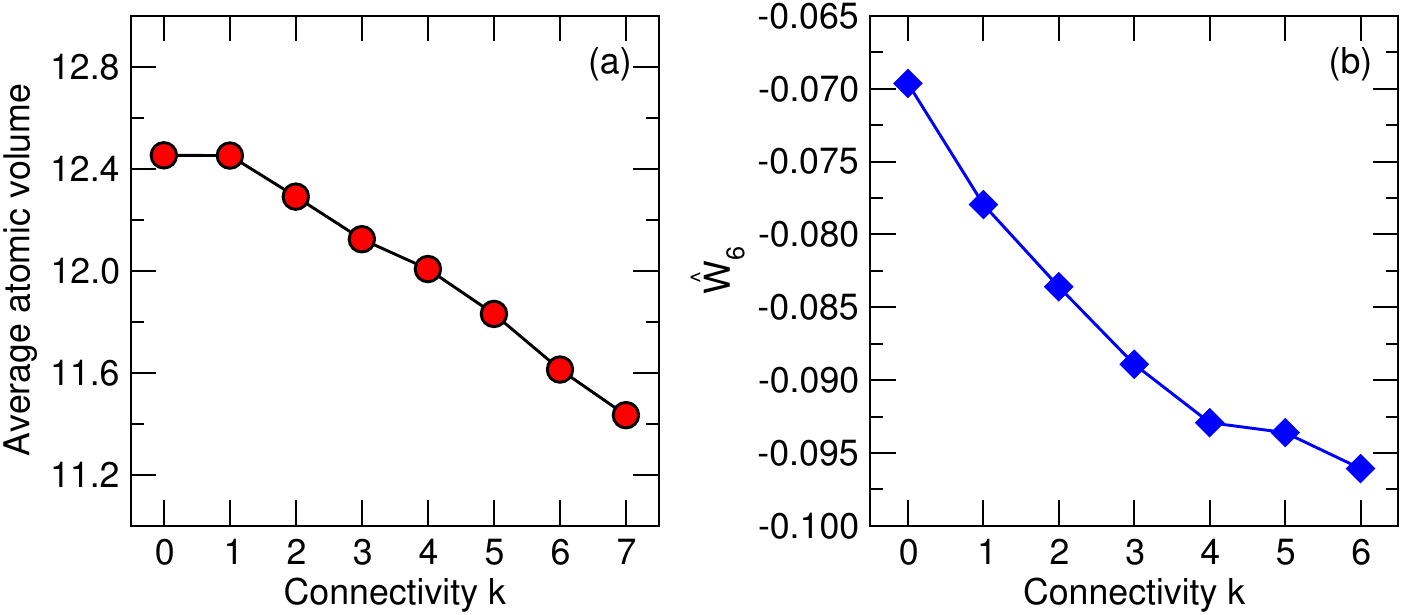}
    \caption{{\bf Local-linking topology integrates the information of ``length'' and ``angle''.} Recent studies have shown that variations in the local connectivity $k$ are linked to corresponding changes in the volume~\cite{wakeda2010icosahedral} and symmetry~\cite{wu2016critical,li2017local} of local icosahedra. As such, for typical metallic liquids connectivity $k$ serves as an order parameter that effectively captures two key physical characteristics, i.e.~{\it length} and {\it angle}, encapsulating changes in both spatial scale and orientational metric. (Adapted from Ref.~\cite{wakeda2010icosahedral,wu2016critical})
    }
    \label{fig9:vsk}
\end{figure}

At the close of this section, we examine the deeper implications of local connectivity. Specifically, why does this parameter, which encodes the local-linking topology of particles, govern such a wide spectrum of physical properties? In solid-state physics, two fundamental quantities often function as order parameters: length and angle. Length, related to either system-wide or local density, commonly acts as the order parameter in first-order phase transitions. Angle, on the other hand, features prominently in modern condensed matter research because it directly corresponds to symmetry or the phase of a quantum wave-function, thereby influencing many physical properties in quantum systems. Interestingly, recent studies demonstrate that both the average atomic volume (associated with length) and the bond-orientational metric $\hat{W}_6$~\cite{Steinhardt1983bond} (linked to angle) exhibit a single-value relationship with connectivity $k$~\cite{wakeda2010icosahedral,wu2016critical}, see Fig.~\ref{fig9:vsk}. This finding suggests that local connectivity can serve as an order parameter that simultaneously incorporates both fundamental structural features--length and angle, and consequently this dual functionality may explain why connectivity $k$ is effective in capturing essential characteristics across various contexts in the case of metallic glass-forming systems.

\section{Topologically non-trivial network and graph-dynamics correspondence}\label{sec4}

In this section, we present a parallel comparative study of the CuZr and NiAl systems. By systematically comparing these two glass-forming alloys, we aim to identify both the common and distinct features in their local structural organization and dynamic relaxation processes. Our analysis reveals that the diversity in particle relaxation behavior is statistically linked to a global topological feature of the transient network formed by the particles in a specific liquid, thereby we introduce the concept of graph-dynamics (GD) correspondence. The GD correspondence elaborates on the potential for identifying more general static properties, such as certain topological invariants or graph-based metrics, that could universally correlate with dynamics. Finally, at the end of this section, we discuss the thermal fluctuations in particle number to ensure that the previously observed relationship between local connectivity and relaxation kinetics cannot simply be attributed to a non-equilibrium (or aging) effect.

\subsection{Parallel Comparative Study of CuZr and NiAl}

As discussed previously, the perspective of atomic clusters has proven highly effective in characterizing the microstructure of amorphous materials and elucidating their distinctive properties. This framework offers a valuable approach to understanding how the macroscopic behavior of amorphous systems is intrinsically linked to their microscopic structures. Extensive research indicates that the morphology and spatial distribution of atomic clusters play an important role in determining the physical properties of these materials, highlighting the importance of analysis of the structure-dynamic relationship to advance our understanding of amorphous systems~\cite{sheng2006atomic,zhou2024toward}. Among the various atomic clusters, icosahedral clusters have attracted particular attention ~\cite{Li2017five,Peng2010effect,Schenk2002icosahedral,Kelton2003first,Luo2004icosahedral,Shen2009icosahedral}. Previous studies indicate that they play a critical role in the structural and dynamical evolution of metallic glass-forming liquids, where their close packing strongly influences the relaxation processes in supercooled liquids, an essential element in understanding both the liquid-to-glass transition and the glass-forming ability of metallic alloys~\cite{cheng2011atomic,cheng2008alloying,cheng2008relationship,ding2014full,hu2015five}. As the temperature approaches the glass transition point, the icosahedral clusters become significantly more abundant, forming a medium-range network structure connected by polyatomic links. This emerging network-like ordering is a hallmark of icosahedral clusters. Furthermore, their geometric properties, such as sharing seven atoms to create a double-topped pentagonal arrangement with fivefold symmetry~\cite{peng2011structural}, have been closely correlated with glass-forming ability, underscoring their importance in understanding the vitrification process~\cite{hu2015five}.

\begin{figure}[ht]
    \centering
    \includegraphics[width=0.9\columnwidth]{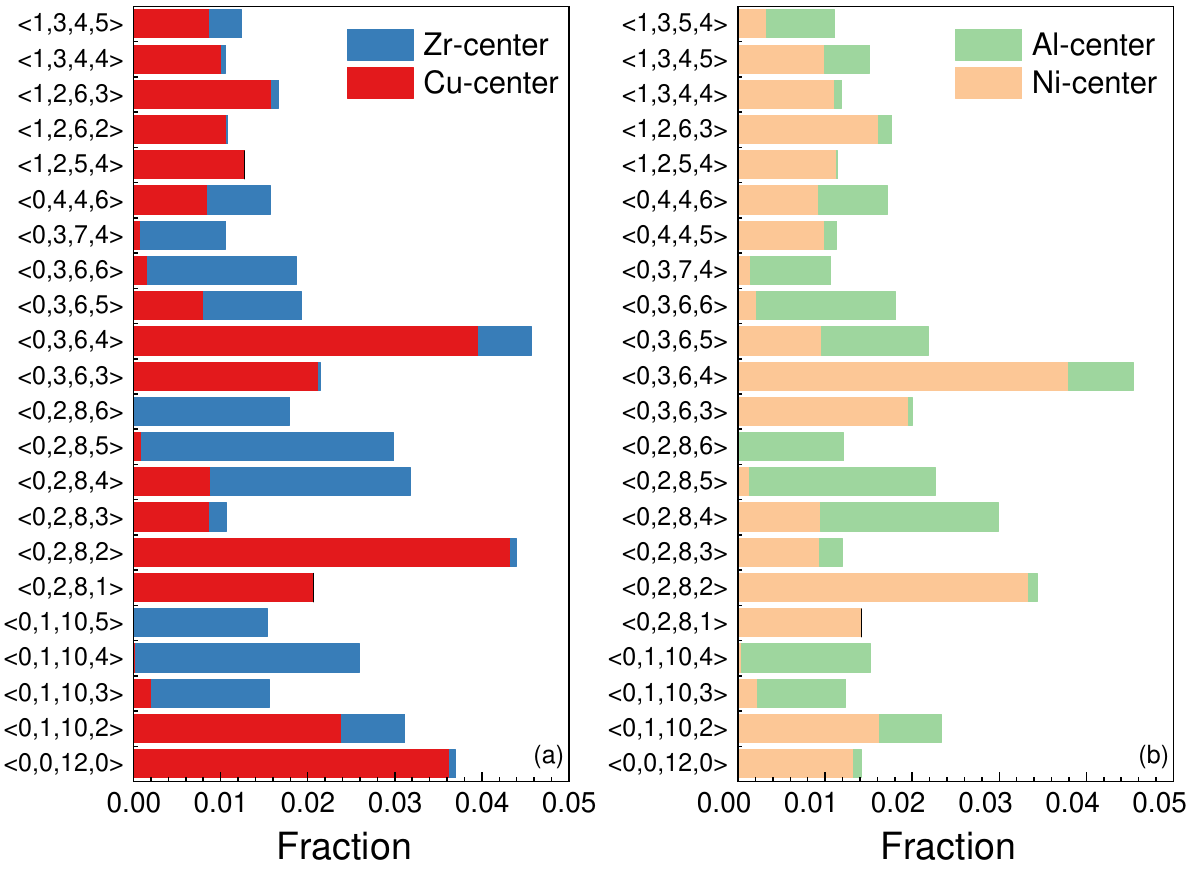}
    \caption{{\bf Voronoi short-range order.} Population of Voronoi polyhedra in liquid Cu$_{50}$Zr$_{50}$ at 1000K (a) and Ni$_{50}$Al$_{50}$ at 900K (b), in which both systems share the same degree of supercooling. One clear difference is that, under similar degrees of supercooling, the fraction of ico atoms is higher in CuZr than in NiAl.
    }
    \label{fig10:voronoi}
\end{figure}

Although icosahedral clusters play a critical role in CuZr metallic liquids structurally and dynamically~\cite{li2017local,liu2013systematic,hao2011dynamic,soklaski2013connectivity}, it remains unclear whether they exhibit similar features in other metallic liquid systems~\cite{li2017local}. It is also uncertain whether properties comparable to those of icosahedral clusters arise in other types of clusters. These questions are central to deepening our understanding of the structural foundations that govern the dynamics of metallic glass-forming liquids. Here we first represent the local environments and dynamical behaviors of several key atomic clusters in the supercooled Cu$_{50}$Zr$_{50}$ liquid, and then apply the same approach to the Ni$_{50}$Al$_{50}$ system. It is found that the specific characteristics of icosahedral clusters are not universal across all metallic liquid systems. Instead, the unique structural and kinetic behaviors of these clusters appear closely tied to the particular liquid metal under study. Next, we will quantify this behavior in more detail and explore its underlying origins in the perspective of complex networks.

Figure~\ref{fig10:voronoi} shows the atomic cluster compositions (i.e., the short-range order) of Cu$_{50}$Zr$_{50}$ and Ni$_{50}$Al$_{50}$. To ensure that both samples have the same degree of supercooling $T/T_\mathrm{m}$, where $T_\mathrm{m}$ is the melting temperature of the corresponding B2 crystal phase (approximately 1723 K for CuZr~\cite{mendelev2007using} and 1535 K for NiAl~\cite{tang2013anomalously}), we selected configurations at $T$=1000K for CuZr and $T$=900K for NiAl, respectively. Using Voronoi polyhedral analysis, we extracted the local atomic structures of both supercooled liquids, as illustrated in Fig.~\ref{fig10:voronoi}. Our results show that the same types of atomic clusters dominate in both systems, and most cluster species overlap between CuZr and NiAl. Notably, Ni$_{50}$Al$_{50}$ lacks clusters of $\langle 0,1,10,5 \rangle$ and $\langle 1,2,6,2 \rangle$ among its major types, whereas Cu$_{50}$Zr$_{50}$ does not contain $\langle 1,3,5,4 \rangle$ and $\langle 0,4,4,5 \rangle$. One distinct difference is that, under comparable supercooling conditions, the fraction of ico atoms is higher in CuZr than in NiAl. Moreover, clusters that appear in higher proportions in NiAl also tend to appear in higher proportions in CuZr. Additionally, Ni-centered clusters in NiAl correspond to Cu-centered clusters in CuZr, while Al-centered clusters in NiAl correspond to Zr-centered clusters in CuZr. These observations confirm a strong resemblance in the local structures of Cu$_{50}$Zr$_{50}$ and Ni$_{50}$Al$_{50}$, providing the rationale for choosing these two systems for comparative analysis. Despite these local structural similarities, further investigation shows that other factors, particularly kinetic behavior, differ significantly between the two systems~\cite{tang2013anomalously}, see below for more details.

\begin{figure*}[ht]
    \centering
    \includegraphics[width=0.85\textwidth]{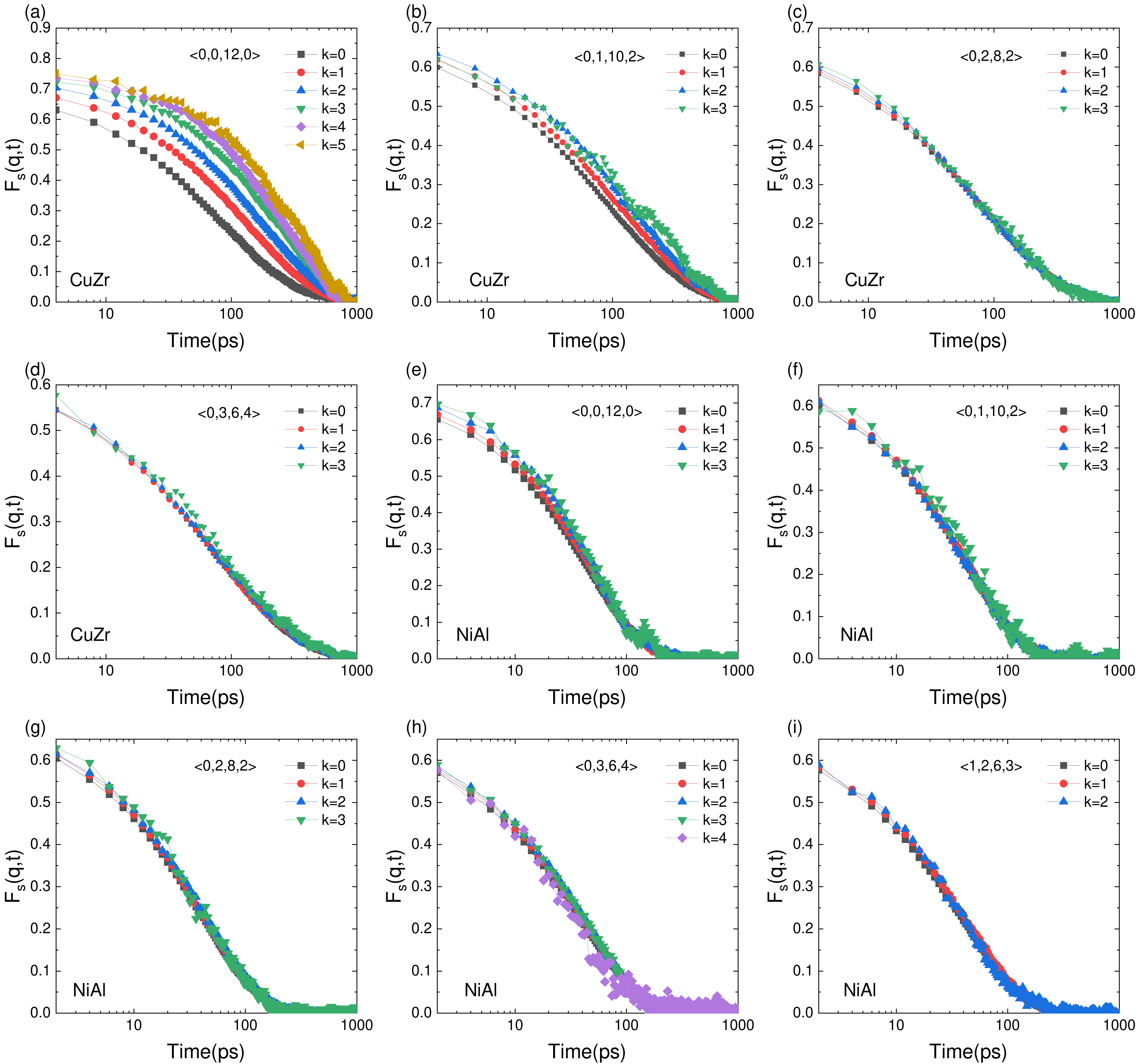}
    \caption{{\bf $k-$dependent relaxation dynamics.} (a)-(d) $F_s(q,t)$ at wave-vector $q=2.8$\AA$^{-1}$ for different values of connectivity $k$ of the Cu atoms coordinated by four different types of Voronoi polyhedron in liquid Cu$_{50}$Zr$_{50}$. $T=1000$K.
    (e)-(i) $F_s(q,t)$ at wave-vector $q=3.04$\AA$^{-1}$ for different values of connectivity $k$ of the Ni atoms coordinated by five different types of Voronoi polyhedron in liquid Ni$_{50}$Al$_{50}$. $T=900$K.
    }
    \label{fig11:ikdyna}
\end{figure*}

\begin{figure}[ht]
    \centering
    \includegraphics[width=0.75\columnwidth]{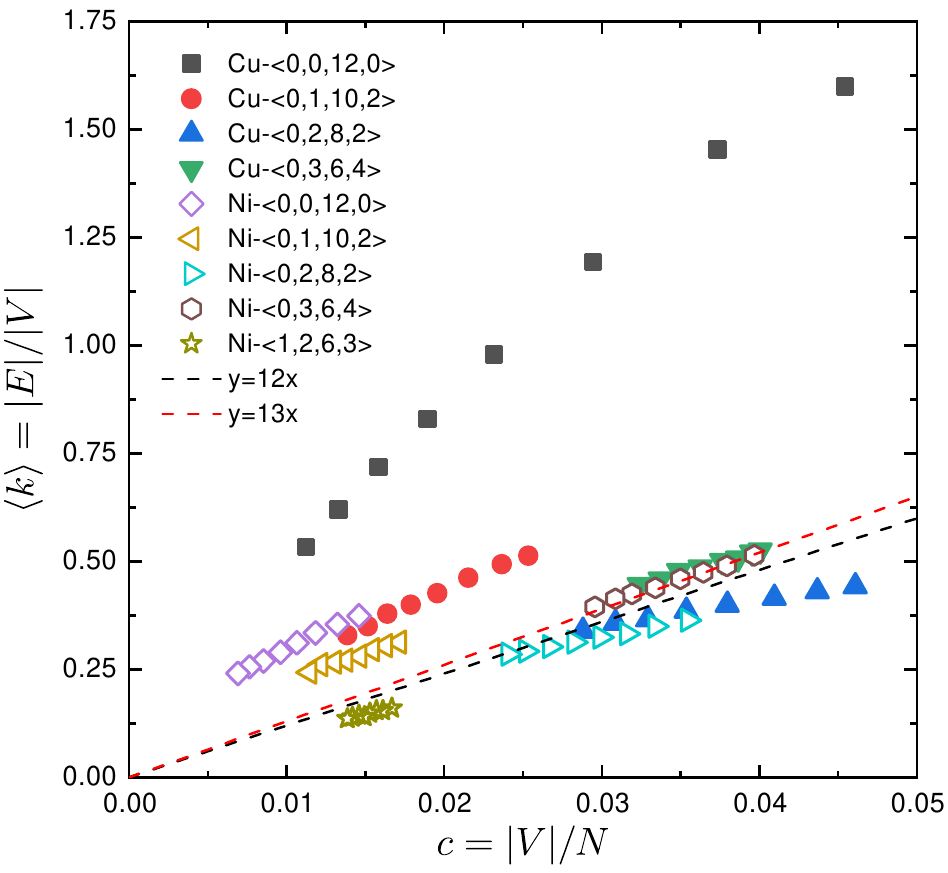}
    \caption{{\bf Complex vs. random graph net.} Average local connectivity $\langle k \rangle = |E|/|V|$ for the major Cu-centered (solid symbols) and Ni-centered (open symbols) atomic cluster types in CuZr and NiAl respectively, plotted as a function of their concentration $c = |V|/N$. The dashed lines indicate the expected average local connectivity from a random distribution of atomic clusters with two different coordination numbers~\cite{li2017local}. The data points cover temperatures ranging from 950K to 1300K, in a step of 50K. $|E|=\# edges$, $|V|=\# nodes$, in sense of graph theory. $N$ is the total number of particles in the MD simulation box.
    }
    \label{fig12:evdiag}
\end{figure}

Figure~\ref{fig11:ikdyna}(a)-(d) present the SISF for the four most common types of atomic clusters in Cu$_{50}$Zr$_{50}$ at 1000K. For clarity, the noisy and scattered SISF of particles with very high connectivity that are present in very small numbers were excluded from the analysis. As shown, all considered cluster types exhibit similar SISF decay trends. However, when examining the effect of connectivity on particle's dynamics, we find distinct patterns. It is revealed that for $\langle 0,2,8,2\rangle$, and $\langle 0,3,6,4\rangle$, the SISF curves calculated at different connectivity $k$ are almost identical, indicating that interconnections do not influence their kinetic behavior very much, which is significantly different from the behavior of the $\langle 0,0,12,0\rangle$ (See Fig.~\ref{fig5:isf_k}, Fig.~\ref{fig6:isf_transport}(a), and Fig.~\ref{fig11:ikdyna}(a)), suggesting a distinctive localization of atoms with icosahedral short-range order in CuZr system. This is consistent with the earlier findings in CuZr liquid with a different chemical element ratio~\cite{li2017local}. Next, we use $\langle k\rangle = \sum k P(k)$~\cite{wu2013correlation}, where $P(k)$ is the probability distribution of connectivity for a given cluster type, to quantify the average local structural features. In line with Li {\it et al.}~\cite{li2017local}, Figure~\ref{fig12:evdiag} illustrates how $\langle k\rangle = |E|/|V|$ varies with the population $|V|/N$ of particles with a specific Voronoi index, i.e. short-range orders, in CuZr supercooled liquid. As the particle population rises, average local connectivity also increases. However, the growth behavior differs among these clusters: $\langle 0,0,12,0\rangle$ exhibits a notably higher average connectivity than the others that is basically in a case of random distribution (i.e. aligning the dashed line as shown in Fig.~\ref{fig12:evdiag}). These observations are consistent with earlier studies~\cite{li2017local} and underscore the unique structural features of $\langle 0,0,12,0\rangle$, which parallel its distinctive kinetic behavior.

The above results demonstrated that icosahedral coordinated atoms in CuZr exhibit distinct structural and kinetic characteristics compared to other type of atoms with a specific local symmetry environment. This raises the question of whether icosahedral clusters in all metallic supercooled liquid systems exhibit such specificity. To address this, we analyzed the structural and kinetic properties of various atomic clusters in Ni$_{50}$Al$_{50}$ using a similar approach. As shown in Fig.~\ref{fig11:ikdyna}(e)-(i), we measured the corresponding SISF for the five most prevalent types of clusters in Ni$_{50}$Al$_{50}$ at $T$=900K. The results reveal similarities in the variation of cluster dynamics between NiAl and CuZr. For example, clusters such as $\langle 0,2,8,2 \rangle$ and $\langle 0,3,6,4 \rangle$ etc. display comparable dynamic behaviors, with similar decay trends in their SISF, suggesting that their dynamics are basically unaffected by the parameter of connectivity in both systems. The notable difference emerges for $\langle 0,0,12,0 \rangle$-centered particles in NiAl. Unlike CuZr, there is also no significant correlation between the SISF of these particles and their local connectivity in NiAl, as evidenced by the overlapping SISF curves for different connectivity $k$. This finding indicates that the specificity of $\langle 0,0,12,0 \rangle$ clusters is significantly diminished in NiAl alloys. Additional measurements were performed at even lower temperatures (the data was not shown here directly), and it indicates that, with only slight effects observed for $\langle 0,0,12,0 \rangle$ (in fact, although the icosahedra-network (ico-network) in NiAl does not diverge from the random network distribution as strongly as in CuZr, it still shows more deviation from random network behavior compared to networks formed by other types of atomic clusters.), for most type of clusters in NiAl, their interconnections had minimal impact on their thermal average kinetic behaviors.

\subsection{Graph-Dynamics correspondence}

Based on previous observations, non-specific clusters—those whose dynamics remain unaffected by their connectivity—exhibit a characteristic linear relationship between the average connectivity $\langle k \rangle$ and their population in terms of local structural features. To validate this hypothesis, we performed simulations of Ni$_{50}$Al$_{50}$ over the temperature range of 950K to 1300K. As shown in Fig.~\ref{fig12:evdiag}, clusters such as $\langle 0,1,10,2 \rangle$, $\langle 0,3,6,4 \rangle$, and $\langle 0,2,8,2 \rangle$ exhibit a consistent linear increase in $\langle k \rangle$ with concentration, which aligns with our findings in the CuZr system. Similarly, $\langle 0,0,12,0 \rangle$ cluster, known for its lack of dynamic specificity, also loses structural specificity in NiAl.

This observation indicates a strong link between dynamical and structural specificity: whenever one is absent, the other is as well. We refer to this interdependence as the {\it graph-dynamics correspondence}, highlighting the fundamental connection between the topological (graph-based) properties of atomic clusters and their dynamic behaviors. This unified perspective offers a powerful framework for analyzing structure–dynamics relationships in supercooled liquids. Notably, evidence suggests that the dynamic behavior of nodes within these networks is closely tied to the sparsity of the overall topology~\cite{ghavasieh2024diversity}, pointing to a deeper interplay between local node dynamics and global structural characteristics. Further investigation into this relationship may provide valuable insights into the mechanisms that drive the development of these topological features. We will come back again to this point in Section~\ref{sec5} (see Fig.~\ref{fig15:GDcorresp}).

Moreover, our results reveal that non-specific clusters behave consistently across different supercooled liquid systems. In NiAl, both the $\langle 0,0,12,0 \rangle$ cluster and other clusters show negligible differences in their dynamic and structural characteristics, suggesting that networking-specificity is not an intrinsic property of atomic clusters with a specific symmetry but rather depends on the system in which they are embedded. A compelling question for future research is whether other clusters with unique specificity exist in NiAl, which could provide further insights into the interplay between structure and dynamics in supercooled liquids.

\vspace{5mm}

\subsection{Thermal fluctuation}

\begin{figure}[ht]
    \centering
    \includegraphics[width=0.85\columnwidth]{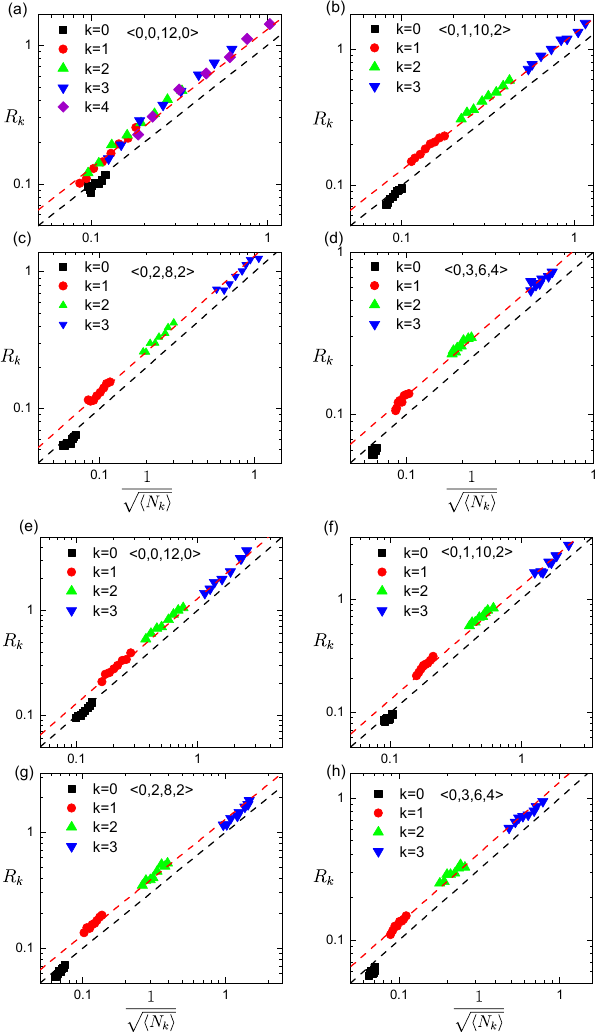}
    \caption{{\bf Particle Number Fluctuations.}
    Relationship between the relative fluctuation $R_k$ and the particle number $N_k$ for clusters with different connectivity $k$ in Cu$_{50}$Zr$_{50}$ (a-d) and Ni$_{50}$Al$_{50}$ (e-h), respectively. The data points for each $k$ corresponds a temperature range of 950K to 1300K. The black dashed line indicates $y=x$, while the red dashed line corresponds to $y=1.3x$.
    }
    \label{fig13:CuZr-NiAl_R_VS_N}
\end{figure}

Although dynamics and thermodynamics describe different facets of a system, they are fundamentally interconnected. In equilibrium, dynamic processes such as thermal fluctuations are dictated by the system’s thermodynamic state. For instance, the fluctuation-dissipation theorem directly links a system’s response to small perturbations with the equilibrium fluctuations predicted by thermodynamics, offering a way to understand how systems relax to equilibrium and how transport coefficients (e.g., viscosity and diffusion constants) relate to thermodynamic quantities. In glassy materials, however, the system departs from equilibrium, and relaxation times grow dramatically as the material approaches vitrification. Thermodynamic properties, such as configurational entropy~\cite{sciortino2005potential}, are closely tied to this dynamic slowdown. Whereas dynamics focuses on how a system changes over time, thermodynamics examines equilibrium properties and state functions that characterize its macroscopic state. Understanding the interplay between these two perspectives is crucial for decoding the behavior of complex systems, particularly supercooled liquids and glasses. In these materials, dynamic phenomena often reveal underlying thermodynamic features, and studying this relationship provides deeper insights into processes like relaxation, aging, phase transitions, and the emergence of non-equilibrium behavior.

\begin{figure}[ht]
    \centering
    \includegraphics[width=0.96\columnwidth]{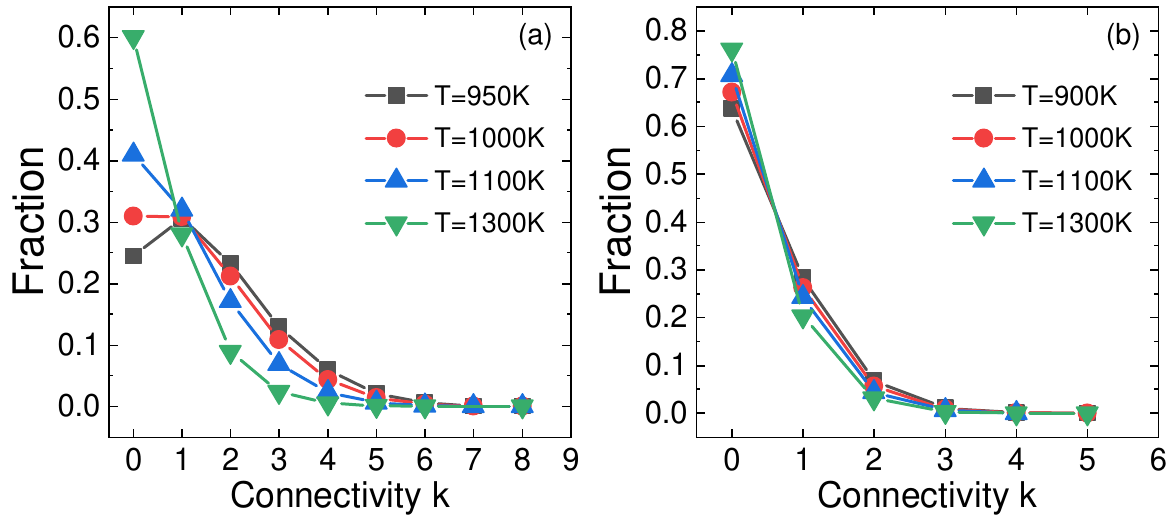}
    \caption{{\bf Probability that an icosahedron ($\langle 0,0,12,0 \rangle$) is of type $k$.} (a) Cu$_{50}$Zr$_{50}$, (b) Ni$_{50}$Al$_{50}$. In CuZr, when temperature decreases, the change in the relative proportions of particles with different $k$ is very pronounced.
    }
    \label{fig14:kdistr}
\end{figure}

Our primary objective here in discussing the physical properties associated with thermodynamic fluctuations is to ensure that the previously observed relationship between atomic connectivity and relaxation kinetics cannot be dismissed as merely a none-quilibrium (or aging) effect. To this end, we investigated the thermodynamic fluctuation behavior~\cite{kubo1973fluctuation,bertini2015macroscopic,bertini2002macroscopic,bernardin2008stationary} of atoms with varying connectivity $k$ in the studied supercooled liquids. Let $N_k$ represent the number of atoms with connectivity $k$ in a sample at temperature $T$. The absolute fluctuation of $N_k$ at this temperature is expressed as $\langle N_k^2 \rangle - \langle N_k \rangle^2$, where $\langle \cdot \rangle$ denotes the ensemble average. Assuming that $M$ configurations were collected at this temperature, the fluctuation is then defined as
\begin{equation}
\resizebox{.905\hsize}{!}{$\langle N_k^2 \rangle - \langle N_k \rangle^2 = \frac{1}{M} \sum_{n=1}^{M} N_k^2(n) - \left( \frac{1}{M} \sum_{n=1}^{M} N_k(n) \right)^2$~,}
\end{equation}
where $N_k(n)$ is the number of atoms with connectivity $k$ in the $n$-th configuration. Then the relative fluctuation of the number of atoms at a specific $k$ is given by $R_k = \sqrt{\frac{\langle N_k^2 \rangle - \langle N_k \rangle^2}{\langle N_k \rangle}}$, and according to textbook knowledge of statistical physics~\cite{landau2013statistical}, for a system in equilibrium, the relative fluctuation of a specific particle type satisfies $R_k \sim 1/\sqrt{\langle N_k \rangle}$. To explore this relationship, we calculated the thermal fluctuation of atoms with different $k$ in the metallic liquids, and the results are shown in Fig.~\ref{fig13:CuZr-NiAl_R_VS_N}(a)-(d) for Cu$_{50}$Zr$_{50}$ and Fig.~\ref{fig13:CuZr-NiAl_R_VS_N}(e)-(h) for Ni$_{50}$Al$_{50}$, respectively.
It can be observed in Fig.~\ref{fig13:CuZr-NiAl_R_VS_N} (a log-log plot) that the relative fluctuations in both systems follow the $\sim 1/\sqrt{\langle N_k \rangle}$ scaling law. Specifically, the data for ``isolated'' atoms (i.e. $k=0$) coordinated by each different Voronoi polyhedron are generally distributed around $y=x$, while of that atoms with $k>0$ are distributed along the $y=1.3x$ line, a hint might suggest that interconnection among atoms would generally suppress the thermal fluctuations. Also, atoms with lower $k$ values display larger fluctuations, indicating their comparatively lower stability. Notably, the icosahedrally ordered Cu atoms (Cu-$\langle 0,0,12,0 \rangle$) in liquid CuZr also follow the $1/\sqrt{\langle N_k \rangle}$ scaling, showing that even atoms with unusual dynamic features exhibit the same thermal fluctuation behavior as other particles. This result confirms that the nontrivial, $k-$dependent dynamics stemming from their distinct linking topology represent a genuine physical phenomenon, rather than a transient non-equilibrium (or aging) effect, paving the way for experimental observation of this effect. Although Cu-$\langle 0,0,12,0 \rangle$ exhibits the same scaling behavior as the other clusters, there are differences. As shown in Fig.~\ref{fig13:CuZr-NiAl_R_VS_N}, the ranges of $R_k$ values for these atoms vary. Across the temperature range considered, the distributions of atoms with different $k$ values generally do not overlap, except for the icosahedrally ordered Cu atoms in liquid CuZr, i.e. in the group of Cu-$\langle 0,0,12,0 \rangle$, the fluctuation ranges of particles with high and low $k$ values overlap. It can be understood by the fact that, for Cu-$\langle 0,0,12,0 \rangle$, the distribution of $k$ at different $T$ showing a significant relative change in the number of particles with various $k$ (see Fig.~\ref{fig14:kdistr}), suggesting a novel developing networking-behavior of ico atoms in CuZr liquid during supercooling~\cite{wu2016critical}.

\section{Perspective and Outlook}\label{sec5}

\begin{figure*}[ht]
    \centering
    \includegraphics[width=0.9\textwidth]{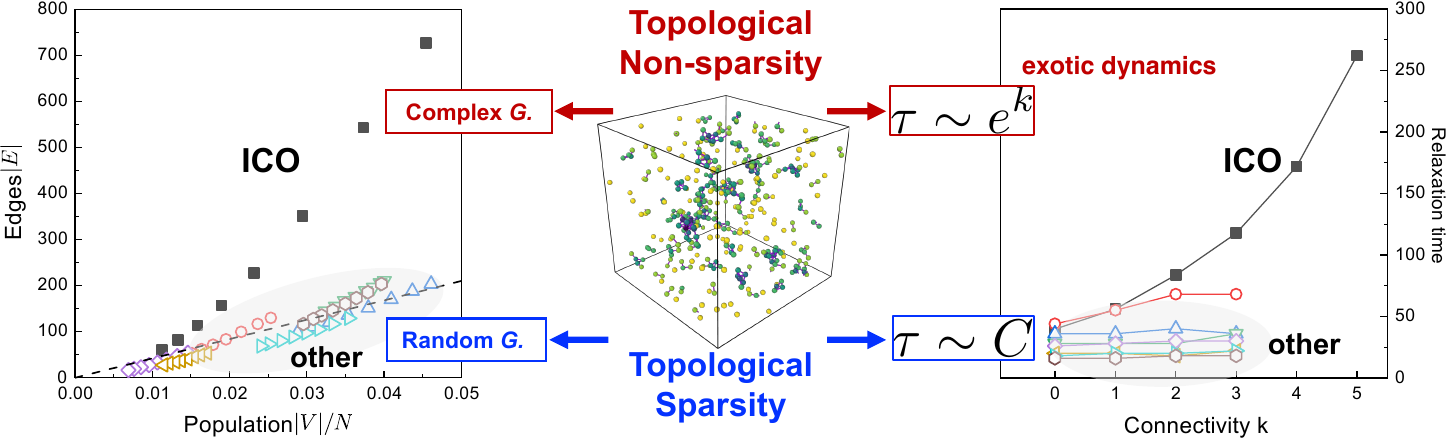}
    \caption{{\bf Graph-Dynamics correspondence.} Recent studies have uncovered significant connections between the emergence of topological features in ico-networks and the local relaxation dynamics and transport behaviors of their constituent components. However, the fundamental origins and nature of some of the most prominent global topological features in ico-networks remain poorly understood. Notably, evidence suggests that the dynamic behavior of nodes within a network is closely linked to the sparsity of its overall topology ($|E| \sim |V|^{\gamma}$: $\gamma$=1, the dashed line in the left panel, is the ideal case~\cite{ghavasieh2024diversity}). This relationship indicates a deeper interplay between local node dynamics and global structural characteristics. Further exploration of this connection could provide valuable insights into the mechanisms governing the evolution of topological features in complex networks.
    }
    \label{fig15:GDcorresp}
\end{figure*}

{\it open question--}
While our results show a strong correlation between local connectivity and particle-level dynamics, particularly for icosahedral atoms, this behavior is not universal for all atomic clusters. It emphasizes that while specific local orders such as icosahedral clusters provide valuable insights, they may not capture the full complexity of glassy dynamics across different systems. The GD correspondence (see Fig.~\ref{fig15:GDcorresp}) reveals significant connections between the emergence of topological features in networks and the local relaxation dynamics and transport behaviors of their constituent components. However, the underlying origins and nature of some of the most prominent global topological features in the ico-networks remain poorly understood. Despite detailed insights from the microscopic mechanisms illustrated in Fig.~\ref{fig9:vsk}, the emergence of the most prevalent global topological features in ico-networks remains unclear. While node connections form through local, microscopic interactions, these underlying rules (although capable of generating global effects that can be quantified by network entropy and free energy) do not directly account for the formation of macroscopic structures.

Recent studies have shown that in real-world networks~\cite{newman2003structure}, the emergence of topological features such as modularity, small-worldness, and heterogeneity aligns with a trade-off between maximizing information exchange and ensuring response diversity over intermediate to large temporal scales~\cite{ghavasieh2024diversity}. In essence, these networks naturally tend to minimize the number of connections between vertices, thereby reducing the associated ``energy cost'', while still maintaining efficient information diffusion~\cite{Busiello2017explorability}. Our observations indicate that particles with connectivity-dependent dynamics tend to form complex graph structures, whereas others yield more random networks. However, the intrinsic nature of these complex ico-networks remains insufficiently understood. Despite notable progress in complex network theory, no well-established synthetic network model currently exists that can replicate the distinctive properties of ico-networks found in certain metallic glass-forming liquids~\cite{albert2002statistical,barthelemy2011spatial,dall2002random}. Developing an ideal model network (or refining classical synthetic network models through subtle modifications) would significantly advance our understanding of atomic networking behavior and the resulting structure-property relationships in glassy materials~\cite{li2024oxidation}. The primary challenge in achieving this lies in our limited geometric understanding of complex networks, even though substantial statistical insights into their overall characteristics have been attained. This open problem presents an exciting avenue for future research.

{\it Graphical Hamiltonian--}
In complex physical systems such as amorphous materials, understanding structure–property relationships is particularly challenging. Identifying the key structural units is non-trivial; these units may be individual atoms, atomic clusters, or even specific dynamic-events. This complexity makes it difficult to define ``effective'' interactions between the elements under study, such as the equivalent interactions between dynamic-events. From a graph theory perspective, however, it is possible to consistently define connections between events or clusters that evolve over time. In constructing graph representations of these complex systems, two core elements emerge: the effective units (vertices), analogous to quasiparticles in condensed matter physics, and the connections (edges) between them, representing generalized interactions. By capturing these local-linking relationships, we can build an adjacency matrix that, when combined with statistical physics methods, facilitates the analysis of structure–property relationships in amorphous materials using a graph-network approach. Notably, this adjacency matrix may, to some extent, function as a ``graphical'' Hamiltonian—a preliminary concept that warrants further investigation.

{\it Real-space Chern number--}
Real-space Chern numbers are a valuable tool for understanding how local structure and disorder impact the global topological characteristics of complex materials and networks. By examining the Chern number directly in real space, rather than relying on momentum space methods, one can reveal how local variations influence overall behavior. In network theory, this approach helps characterize how topological properties shape the flow, connectivity, and dynamics of complex systems. Techniques like the Kitaev lattice curvature formula allow for evaluating a local Chern number by summing contributions within a finite region~\cite{mitchell2018amorphous,kitaev2006anyons}. Dividing the network into clusters and calculating a Chern number for each provides an approximate global value. In highly disordered systems, local Chern numbers may differ significantly from place to place, making a careful handle of the calculation necessary to ensure a meaningful global characterization of topological features across heterogeneous and complex structures.

{\it Concluding remarks--}
Topology, a well-established branch of mathematics, classifies objects based on invariant geometric properties and has significantly advanced our understanding of fields such as condensed matter physics, cosmology, and particle physics. However, its application in glass physics and materials science remains underdeveloped. In this work, we highlight the need to move beyond conventional approaches to address the formidable challenges posed by the highly complex, multiscale structural and dynamical behaviors in this field. We hope that our findings will inspire new insights and encourage prospective researchers to actively engage in advancing this promising direction.

\vspace{5mm}

\noindent
X.-J. Zhou, F. Yang, X.-D. Yang, and L. Ma contribute equally to this work.

\vspace{5mm}

\noindent
{\bf Acknowledgments}
~Z.W.W. would like to express his sincere gratitude to all the collaborators who have made significant contributions to the research on this topic, as well as to the many colleagues whose ideas and insightful discussions have greatly benefited the relevant study. This work was supported by the National Natural Science Foundation of China (Grant Nos. 12474184, 52031016, and 11804027).

\bibliography{ref} 

\end{document}